\def\year{2021}\relax
\documentclass[letterpaper]{article} 
\usepackage{aaai21} 
\usepackage{times} 
\usepackage{helvet} 
\usepackage{courier} 
\usepackage[hyphens]{url} 
\usepackage{graphicx} 
\usepackage{graphicx} 
\usepackage{natbib} 
\usepackage{caption} 
\usepackage{amssymb}
\usepackage{mathtools}
\usepackage[binary-units=true]{siunitx}
\usepackage{subcaption}
\usepackage{algorithm}
\usepackage{algorithmicx}
\usepackage[noend]{algpseudocode}
\usepackage{amsthm}
\usepackage{xcolor}

\newcommand{\ignore}[1]{}

\title{Multi-Target Search in Euclidean Space with Ray Shooting \\ (Full Version)}
\author{Ryan Hechenberger, Daniel Harabor, Muhammad Aamir Cheema,\\Peter J Stuckey, Pierre Le Bodic\\
}
\affiliations{
Faculty of Information Technology, Monash University, Australia\\
\{Ryan.Hechenberger, Daniel.Harabor, Aamir.Cheema, Peter.Stuckey, Pierre.LeBodic\}@monash.edu
}
\graphicspath{ {img/} }

\newcommand{\Pt}[1]{{#1}}
\newcommand{\Ptd}[2]{\vec{#1#2}}
\newcommand{\Pdist}[2]{\|\vec{#1}{#2}\|}

\newcommand{\CW}{\textsf{cw}}
\newcommand{\CCW}{\textsf{ccw}}
\newcommand{\Start}{\ensuremath{s}}
\newcommand{\StartPt}{\ensuremath{\Pt{\Start}}}
\newcommand{\Target}{\ensuremath{t}}
\newcommand{\TargetPt}{\ensuremath{\Pt{\Target}}}

\newcommand{\Ffn}[1]{\operatorname{f}(#1)}
\newcommand{\Gfn}[1]{\operatorname{g}(#1)}
\newcommand{\Hfn}[1]{\operatorname{h}(#1)}
\newcommand{\AS}[2]{\operatorname{AS}(#1,#2)}
\newcommand{\ASsplit}[3]{\operatorname{SPLIT}(#1,#2,#3)}

\newcommand{\true}{\mathit{true}}
\newcommand{\false}{\mathit{false}}

\newcommand{\orayscan}{\textsc{RayScan}}
\newcommand{\rayscan}{\textsc{RayScan}$^+$}

\algrenewcommand\algorithmicindent{1.0em}
\newcommand\blfootnote[1]{%
  \begingroup
  \renewcommand\thefootnote{}\footnote{#1}%
  \addtocounter{footnote}{-1}%
  \endgroup
}
\nocopyright
\begin{document}

\maketitle

\begin{abstract}
The Euclidean shortest path problem (ESPP) is a well studied problem with many practical applications.  
Recently a new efficient online approach to this problem, \orayscan{}, has been developed, based on ray shooting and polygon scanning. In this paper we show how we can improve \orayscan{} by carefully reasoning about polygon scans.
We also look into how \orayscan{} could be applied in the single-source multi-target scenario, where logic during scanning is used to reduce the number of rays shots required.
This improvement also helps in the single target case.
We compare the improved \rayscan{} against the state-of-the-art ESPP algorithm, illustrating the situations where it is better.
\end{abstract}

\blfootnote{This is the full version of an extended abstract released The 14th Annual Symposium on Combinatorial Search (SOCS 2021).}

\section{Introduction}

Euclidean shortest path finding (ESPP) has many variants. We examine obstacle-avoiding 2D shortest (i.e. optimal) path determination, which is a well-studied problem with many practical applications, in e.g. robotics and computer games \citep{PathfindingStudy}.
Obstacles are an effective way to represent the world directly, for example, a wall can be represented as a rectangular obstacle that has to be navigated around.

ESPP algorithms can be very fast in \emph{static} environments, i.e. where obstacles do not change between shortest path queries, as this allows for longer pre-processing.
Search queries in a \emph{dynamic} environment are considerably harder, as the obstacles change regularly, therefore any pre-processing has to be updated to account for these changes.

Most existing approaches to ESPP first convert obstacles into another representation to perform a search, either via a visibility graph~\citep{VisGraph} or a navigation mesh~\citep{polyanya} for example. A recent approach is \orayscan{}~\citep{rayscan}.
It partially builds a visibility graph on the fly by using ray shooting (also called ray \emph{casting}) to discover blocking obstacles, and then scanning along the edges of such obstacles to find ways around.
\orayscan{} is comparable to an on-the-fly partial edge generation of a sparse visibility graph (SVG) running taut A* \citep{SparceVG}; as such it does not require pre-processing, as it makes use of the natural polygonal representation of obstacles.
Much of the runtime of \orayscan{} is taken up by the ray shooting.

In this paper we extend \orayscan{} with various improvements that are aimed at reducing the number of ray shots required. We call this advancement \rayscan{}. We also tackle the multi-target variant, where we find the shortest path from a start point $s$ to all target points $t \in T$.

We compare \rayscan{} against the state-of-the-art ESPP algorithm Polyanya \citep{polyanya}; including tests with static and dynamic environments, single- and multi-target search. We show \rayscan{} significantly improves upon
\orayscan{} and that \rayscan{} is competitive with Polyanya (in particular in highly dynamic scenarios).

\begin{figure}[tb]
\centering
\includegraphics[width=0.55\linewidth]{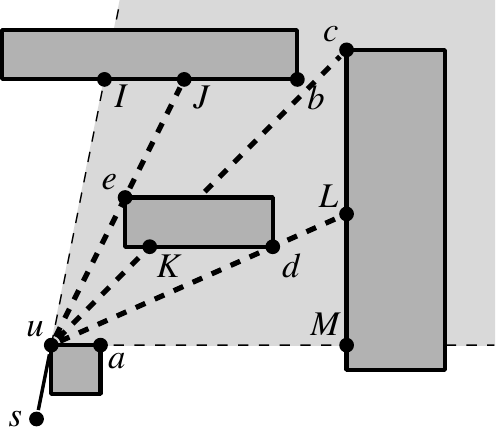}
\caption{Expanding of node $u$}\label{fig:nodeExp}
\end{figure}

\begin{figure*}[t]
\centering
\begin{subfigure}[t]{0.22\textwidth}
\includegraphics[width=.95\columnwidth]{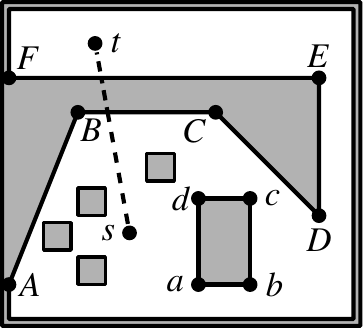}
\subcaption{shoot target}\label{fig:exampleA}
\end{subfigure}
\begin{subfigure}[t]{0.22\textwidth}
\includegraphics[width=.95\columnwidth]{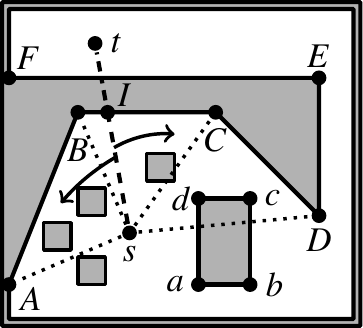}
\subcaption{scan \CCW{} to A, then \CW{} to D}\label{fig:exampleB}
\end{subfigure}
\begin{subfigure}[t]{0.22\textwidth}
\includegraphics[width=.95\columnwidth]{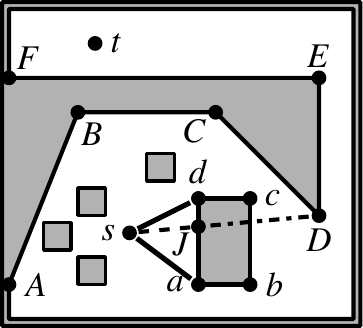}
\subcaption{shoot D and scan intersection}\label{fig:exampleC}
\end{subfigure}
\begin{subfigure}[t]{0.22\textwidth}
\includegraphics[width=.95\columnwidth]{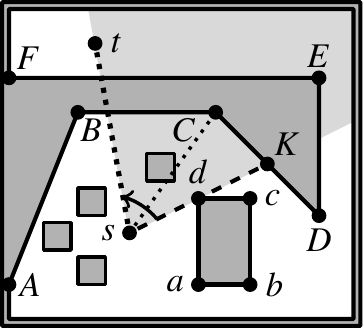}
\subcaption{d is successor, scan within angle sector}\label{fig:exampleD}
\end{subfigure}
\\
\begin{subfigure}[t]{0.22\textwidth}
\includegraphics[width=.95\columnwidth]{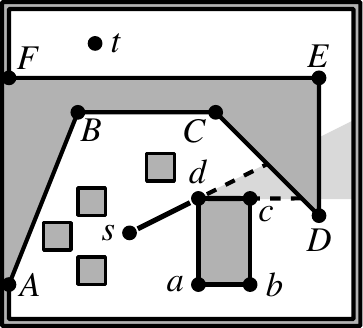}
\subcaption{expand d, c is successor}\label{fig:exampleE}
\end{subfigure}
\begin{subfigure}[t]{0.22\textwidth}
\includegraphics[width=.95\columnwidth]{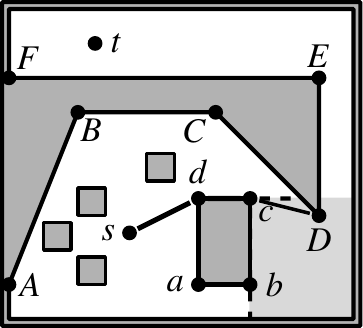}
\subcaption{expand c, D is successor}\label{fig:exampleF}
\end{subfigure}
\begin{subfigure}[t]{0.22\textwidth}
\includegraphics[width=.95\columnwidth]{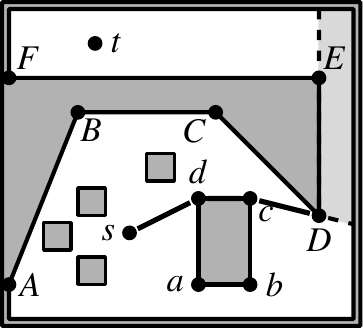}
\subcaption{expand D, E is successor}\label{fig:exampleG}
\end{subfigure}
\begin{subfigure}[t]{0.22\textwidth}
\includegraphics[width=.95\columnwidth]{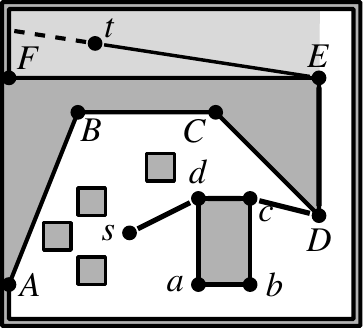}
\subcaption{expand E, t is visible}\label{fig:exampleH}
\end{subfigure}
\caption{\orayscan{} Example~\cite{rayscan} - A dashed line is a ray shot. Each corner scanned by \orayscan{} is shown by a dotted line from $s$ to the corner. The solid lines show the shortest path.}
\label{fig:example}
\end{figure*}

\section{RayScan}
\orayscan{} is presented in this paper as a 2D ESP algorithm, made to work with a 2D environment represented
as a set of non-intersecting polygonal obstacles (\textbf{inner-obstacles}), and a single enclosing polygon
(\textbf{enclosure} or \textbf{outer-obstacle}), containing and non-intersecting all inner-obstacles.
The \textbf{outer-boundary} is defined as the convex hull of the enclosure.

\orayscan{}{} does not provide the method of searching, rather it can be viewed as a fast method of producing a
subset of edges of the visibility graph during the search. It uses the A* algorithm \citep{Astar} to drive the search,
using \textbf{start} ($\StartPt$), \textbf{target} ($\TargetPt$) and corner points on the polygons as the nodes
in the search. When pushing a successor edge $\Ptd{u}{v}$, the edge weight is the Euclidean distance from $u$ to $v$,
$|\Ptd{u}{v}|$.

During the search, we expand nodes by producing successors.
The search is directed with the heuristic function $\Hfn{v} = |\Ptd{v}{t}|$. We use the minimum $f$~value to select the next
node to expand, where $\Ffn{v} = \Gfn{v} + \Hfn{v}$ and $\Gfn{v}$ is the $g$~value (length of the shortest path from $s$ to $v$).

The search is directed by ray shooting, where a ray is shot from the expanding node $u$ along a direction
vector, which returns the first polygon hit by the ray and its intersecting point.
This gives us an obstructing obstacle that we need to navigate around, which in a 2D environment, is restricted to two ways, clockwise (\CW{}) or counter-clockwise (\CCW{}).

The approach to navigate around an obstacle is the \textbf{scan} routine, where we trace a scan~line along the polygon in a
\CW{} or \CCW{} orientation w.r.t. expanding node $u$, e.g., Figure~\ref{fig:nodeExp} shoots ray from  $u$ to $c$, finding intersection $K$; a \CW{}-scan starts from the scan line $uK$ and has it sweep in a \CW{} direction along polygon edge to point $d$ and then stop, finding the $d$ point which we refer to as a \textbf{turning point}.
The \CCW{}-scan has the scan line reach turning point $e$. Determining which point is considered a turning point can be done with two different approaches: the forward- and convex-scanning methods; 
that are detailed under the Scanning section. If there are no
obstacles between $u$ and a turning point, it is considered a \textbf{visible point} and is
a successor node of $u$, otherwise it is a \textbf{blocked point}.

The scanning process does not end at finding the turning point; it instead recurses into new scans.
Referring to Figure~\ref{fig:nodeExp}, for a blocked point like $c$, the scan will split into two scans, one \CW{} the other \CCW{} starting at the intersection point $K$.
For a visible point like $d$ (found from intersection $K$), we add $d$ as a successor and shoot the ray past the point to find the next polygon up (intersection $L$); then we continue the scan in the same orientation (\CW{}) from that intersection ($L$).
The whole recursive process starts from a single scan, where we call the whole process the \textbf{full scan}.

The recursive scanning process makes use of angled sectors to improve performance and as a base condition of the recursion.
An angled sector is depicted as $\AS{\vec{a_\CCW}}{\vec{a_\CW}}$, where the angled sector is defined as the angular region starting from the \CCW{} pivot angle $\vec{a_\CCW}$, turning \CW{} towards angle $\vec{a_\CW}$.
Referring to Figure~\ref{fig:nodeExp}, the shaded area circular to $uI$ and $uM$ is an angled sector ($\AS{\vec{uI}}{\vec{uM}}$).
During a scan, the scan line must always remain within this area; if at any point the scan~line leaves the sector, the recursion ends with no (additional) turning point discovered.  For the start point we abuse notation and use a \SI{360}{\degree} angled sector $\AS{\vec{st}}{\vec{st}}$.

An angled sector can also be split into two angled sectors.
We use the notation $\ASsplit{\AS{\vec{a_\CCW}}{\vec{a_\CW}}}{\vec{p}}{d}$ to split the angled sector $\AS{\vec{a_\CCW}}{\vec{a_\CW}}$ along $\vec{p}$, returning a $d$-split (\CCW{}-split or \CW{}-split). The \CCW{}-split returns $\AS{\vec{a_\CCW}}{\vec{p}}$, while a \CW{}-split returns $\AS{\vec{p}}{\vec{a_\CW}}$.
This split is used by the recursive scanning, where for turning point $p$, any recursive \CCW{} scan will take the \CCW{}-split of the current angled sector along $p$, and vice versa for \CW{} scan.

Every expanding node $u$ (excluding $s$) has a special angled sector known as the projection field.
This angled sector encompasses the region that all successors of $u$ must fall within, since any shortest path via $u$ must be taut to the obstacle that $u$ is part of. In
Figure~\ref{fig:nodeExp}, the projection field for $u$ is $\AS{\vec{su}}{\vec{ua}} \equiv \AS{\vec{uI}}{\vec{uM}}$.
The full scan begins with the projection field, resulting in significant speedups by limiting the area the scan runs in.

A point on a polygon is considered to be a \textbf{concave point} if its inside angle is more than \SI{180}{\degree}, otherwise is a \textbf{convex point}. Expanded nodes, with the possible exception of
$\StartPt$, will always be convex, as you can only bend around such points.

\section{RayScan Example}

Consider the example execution of \orayscan{} illustrated in Figure~\ref{fig:example} from~\cite{rayscan}. Initially we expand $s$ by 
shooting a ray from start $s$ to target $t$ (Figure~\ref{fig:exampleA}). We find this is blocked
by a polygon (the outer polygon). We then recursively scan to move around the blocking polygon from $I$ (Figure~\ref{fig:exampleB}). 
Scanning \CCW{} we skip $B$ since the ray cannot bend around the polygon here
and reach node $A$ which is on the convex boundary of the map, so we stop the scan. There is no way around the (outer) polygon in this direction. Scanning \CW{} we skip $C$ and reach $D$.  We then shoot a ray towards $D$, which is blocked by the polygon $abcd$ at intersection $J$ (Figure~\ref{fig:exampleC}). We recursively scan this polygon. The \CW{} scan will find $a$ as a turning point, shooting to it reveals it is visible and adds to the queue, and shooting past it will hit the outer-boundary, thus the \CW{} scan does not continue. The recursion in the other orientation, \CCW{}, finds turning point $d$, shooting to it and adding it to the queue (Figure~\ref{fig:exampleD}). We shoot past $d$ and find intersection $K$, scan \CCW{} from here we leave the angled sector boundary $\Ptd{s}{t}$, ending the recursion.

Next we expand $d$ (Figure~\ref{fig:exampleE}). The target is not within  the \emph{projection field}
$AS(\Ptd{s}{d},\Ptd{d}{c})$ so we scan both extreme edges. We shoot a ray $\Ptd{d}{c}$ and find that $c$ is visible from $d$, thus  $c$ is added as its successor. We scan the polygon that blocks $\Ptd{d}{c}$ in \CCW{} and leave the angled sector without discovering any turning point and stop. The other extreme ray $\Ptd{s}{d}$ does not discover any successor.
Next, we expand $c$ in the same manner as $d$ (Figure~\ref{fig:exampleF}), scanning \CW{} to find only $D$. The next lowest $f$~value is $a$, which we expand to find $b$ (not shown in figure). Node $b$ is then expanded (not shown in figure) and it finds $D$, although a shorter path to $D$ is already discovered thus it is not added.
Now, we expand $D$ and find $E$ scanning \CW{} (Figure~\ref{fig:exampleG}).

Finally, we expand $E$ and notice that $t$ is within the projection field (Figure~\ref{fig:exampleH}), thus we shoot to $t$. We find that $t$ is visible from $E$,  thus we have found a path to $t$. This path is the shortest path because the node $E$ had the minimum $f$~value which guarantees that all other paths to $t$ are no shorter than this path.

\begin{figure*}[ht]
\centering
\begin{subfigure}[b]{0.2\textwidth}
\includegraphics[width=0.95\linewidth]{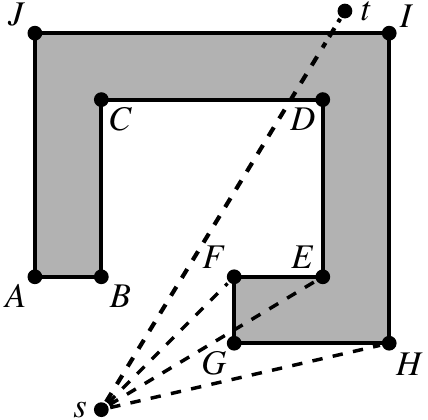}
\subcaption{Scanning \CW{} from $\Ptd{s}{t}$.}\label{fig:scanMethod}
\end{subfigure}
\begin{subfigure}[b]{0.22\textwidth}
\includegraphics[width=0.95\linewidth]{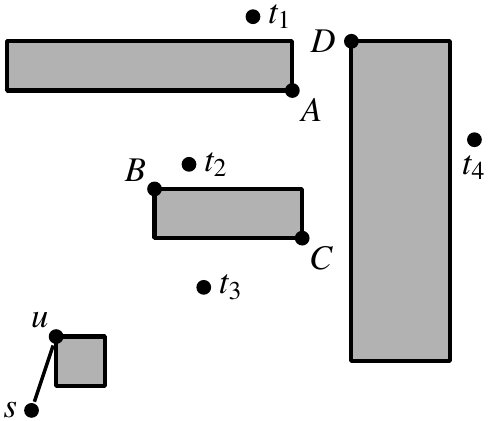}
\subcaption{Blocking extension}\label{fig:blocking}
\end{subfigure}
\begin{subfigure}[b]{0.22\textwidth}
\includegraphics[width=0.95\linewidth]{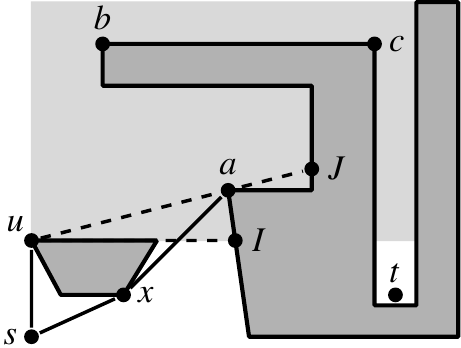}
\subcaption{Skip extension}\label{fig:skip}
\end{subfigure}
\begin{subfigure}[b]{0.22\textwidth}
\includegraphics[width=0.95\linewidth]{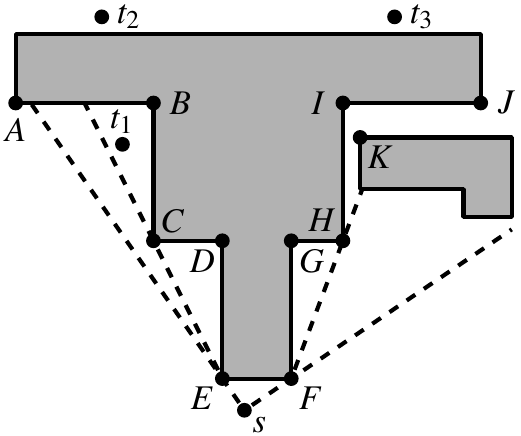}
\subcaption{Bypass extension}\label{fig:bypass}
\end{subfigure}
\caption{Auxilliary diagrams}\label{fig:extensions}
\end{figure*}

\section{RayScan Improvements}

The first contribution is a number of improvements to \orayscan{}.
They do not change the underlying algorithm, but improve components hence the proof of optimality of \orayscan{} continues to apply.  

\subsection{Scan Overlap}
The expansion of a node in \orayscan{} relies on multiple full scans, two from both ends of the projection field, called the \textbf{projection scans} (for any expanding node not $s$), and an additional two if the target is within the projection field, called the \textbf{target scans}.

These full scans usually overlap with each other, resulting in redundant work of up to three complete scans of 
the region. We introduce two methods for reducing this overlap.

\textbf{Refinement by sector:}
The first method, implemented in \orayscan{}, \emph{refinement by sector}, works by refining the initial angle sector given to the full scan. Normally, the angled sector for the full scan is the projection field for projection scans, and the projection field split by $\Ptd{u}{t}$ (where $u$ is expanding node) for target scans.
When we do a full scan that will overlap a previous scan for that expansion, work will become redundant when the scan line passes a ray shot by a previous full scan. To avoid this, we want to identify the closest \CW{} or \CCW{} to the starting scan line and split the angled sector to prevent the scan from passing it.

For example, in Figure~\ref{fig:nodeExp}, starting with a projection scan \CW{} from $\Ptd{s}{u}$, it starts with the angled sector of the projection field $\AS{\Ptd{s}{u}}{\Ptd{u}{a}}$, shoots the rays from $u$ along $\Ptd{s}{u}$, $\Ptd{u}{b}$, $\Ptd{u}{e}$ and $\Ptd{u}{d}$. The next full scan starting \CCW{} from $\Ptd{u}{a}$ will have a reduced angled sector of the projection field $\AS{\Ptd{u}{d}}{\Ptd{u}{a}}$.

\textbf{Refinement by ray:}
The second method, implemented in \rayscan{}, \emph{refinement by ray}, uses the projection field for all full scans, instead ending a scan early when we try shooting a ray towards a turning point that we have already shot to this expansion.

For example, in Figure~\ref{fig:nodeExp}, as in the  previous example we perform expansion of the first projection scan firing rays $\Ptd{s}{u}$, $\Ptd{u}{b}$, $\Ptd{u}{e}$ and $\Ptd{u}{d}$, out of these we shot rays towards turning points $b$, $e$ and $d$. The full scan from $\Ptd{u}{a}$ will find point $c$, shoot towards it and then recurse to find turning points $e$ and $d$, which we have already shot rays to from a previous scan this expansion, thus both scans will end.

\subsection{Scanning}

\textbf{Forward scan:}
The scan is split into two types, \CW{} and \CCW{}. The original \orayscan{} method uses a scan line that rotates in the specified orientation until the line is forced to turn the opposite way when encountering a polygon vertex. This is a turning point. We call this the \emph{forward scan} method.

The idea of a turning point is to find a possible point that is visible from the expanding node and also on the shortest path. The forward scan method can find a concave point, which 
can never be on a taut path and hence not on any shortest path. We also know that a concave turning point can never be visible due to the nature of the scan, if the scan line goes from \CCW{} to \CW{} and is concave, the progressing line must be blocking the point.

For example, consider Figure~\ref{fig:scanMethod}, point $C$, $D$ and $E$ are concave points, when performing a \CW{} scan from ray $\Ptd{s}{t}$, a forward scan will stop at $E$ (as $E$ to $F$ orientates \CCW{}) and shoot a ray. We see that $E$ is blocked by its own polygon, though it will find $F$ and $H$ after the recursive scan.

\textbf{Convex scan:}
We consider an alternate scanning method we call \emph{convex scan}, 
where we only shoot at convex points, which are the only 
points that can appear on a shortest path. Referring back to the previous example, the scan line will instead pass over $E$ even though it goes \textbf{opposite} to the scanning orientation, as $E$ is concave, instead resulting in finding $F$ as a turning point.

Normally when we find a turning point (in this case $F$), and it is a visible turning point, we will shoot the ray past that point and recurse the scan in the same orientation. For the example \CW{} scan we will normally then perform another \CW{} scan, 
though this will simply reach $F$ again. 
We can see that $F$ is a turning point in the opposite orientation to the scan direction.  We refer to this kind of point as a \textbf{backward turning point}. Given a visible-backward turning point we must recursively scan in the opposite orientation \emph{and} continue the scan past $F$. For a \CCW{} scan from $\Ptd{s}{F}$ ray (will go past the $\Ptd{s}{t}$ angled sector) and a \CW{} scan will continue (go past $G$ and end on $H$). This \CW{} scan finds $H$, which is a \textbf{forward turning point}, and is handled as normal.

\subsection{Avoiding Ray Shooting}
\orayscan{} can be improved by avoiding shooting rays that are unnecessary. We list three different methods.

\textbf{Blocking extension:}
Originally, \orayscan{} starts expanding a node  $u$ with the target scans (if $t$ is within $u$'s projection field), followed by the projection scans. It is possible to instead start with the projection scans followed by target scans.
The \emph{blocking extension} makes use of this switch in the order of the full scans to potentially avoid some target scans entirely. If during the projection scans, we see that the target is not visible (i.e. is blocked by an obstacle edge we scanned over), then we do not have to shoot the target.

For example, in Figure~\ref{fig:blocking}, we show several possible targets $t_{\{1,2,3\}}$. Expanding node $u$ will shoot ray $\Ptd{s}{u}$ from $\Pt{u}$ and do a \CW{} scan to find $A$. During that scan $t_1$ is passed by the scan line behind the obstacle scanned, meaning we do not need to shoot to it later. After shooting to $A$, a \CCW{} scan to find $B$, sees that $t_2$ is blocked, while a \CW{} scan will find $C$ and pass over $t_3$, except in this case $t_3$ is in front of the obstacle so we still need to shoot towards it.


\textbf{Skip extension:}
The \emph{skip extension} attempts to skip (scan) past a turning point that cannot be on the shortest path. We can deduce a skip candidate point $v$ from $u$ if the following condition holds:
$
    \Gfn{u} + \Pdist{u}{v} > \Gfn{v}. \label{eq:skipCandidate}
$
Formally, expanding $u$, when finding a turning point $v$ that has been reached by another point $x$, then $\Gfn{v} = \Gfn{x} + \Pdist{x}{v}$, and if the $g$~value $u$ will give $v$ as a successor $\Gfn{u} + \Pdist{u}{v}$ is larger than $v$'s current $g$~value, than this point may be skipped, as we know that candidate point $v$ is not on a shortest path via $u$.
But we may need to find a shortest path to a point past $v$ so the scan may need to continue.

For example, referring to Figure~\ref{fig:skip}, the shortest path is $s-u-b-c-t$. When expanding $u$, we perform a \CCW{}-projection scan that starts at intersection $I$, then finds turning point $a$.
We normally shoot to $a$ and continue the \CCW{} scan from $J$, except in this case we notice $a$ has a shorter path from $s$ via $x$. Since we do not want to reach $a$ from $u$ but need to continue the scan to find the successor on the shortest path $b$, we want to avoid shooting the ray to $a$.
Instead, we continue the scan without shooting, ignoring all turning points (if any).
If the scan line orientation reaches what would be the ray ($\Ptd{u}{a})$ when reaching $J$ without leaving the angled sector (as is the case in this example) then we deduce that we do no need to shoot to $a$ and can continue the \CCW{} scan from $J$ to eventually find the turning point $b$.
Unlike this example, if we are unable to pass ray $\Ptd{u}{a}$, either from leaving the angled sector, reaching $a$ again or encountering a certain number of points (a limit for performance reasons), then we \emph{must} shoot to $a$ to continue the scan.
In any case, the scan must be continued \CCW{} from $J$ as $b$ is on the shortest path from $u$. 

\textbf{Bypass extension:}
The final extension outlined in this paper is the \emph{bypass extension}, which is similar to the skip extension except it can be applied to turning points that are not skip candidates, which do not appear that often.
Bypass seeks to skip shooting a ray to a node in the same way as the skip extension, except it also checks if the scan line sweeps over the target, and if it does not then we can bypass the node.
For example; with Figure~\ref{fig:bypass}, when expanding $E$, the \CCW{} projection scan reaches $C$ and attempts to bypass it.
The scan line will continue from $C$ past $B$ before reaching the ray that would be $\Ptd{E}{C}$; if $t_1$ is our desired target then the scan line sweeps over it, meaning we cannot bypass $C$ (as it is on the shortest path); if however $t_2$ is our desired target; although we sweep past it, we did not sweep over $t_2$ as it is blocked by the polygon, thus we can bypass $C$, which is done the same way as skip.  The scan then immediately leaves the angled sector producing no successors.

Sometimes a node on the shortest path will be bypassed when expanding another node on the shortest path; take Figure~\ref{fig:bypass} as an example, when expanding node $F$; we perform a \CW{} projection scan that finds turning point $H$ (on the shortest path), except the point can be bypassed, thus \rayscan{} will skip over it to find turning point $J$.
This is where considering the angled sector is important, as that bypass was done with the projection field $\AS{\Ptd{F}{G}}{\Ptd{s}{F}}$; however, since we did not shoot a ray to $H$ that angled sector remains the same, this is important as it allows us to reach $H$ further in the full scan. After find turning point $J$ (on the shortest path), we shoot a ray and find it blocked, thus we recurse, the \CCW{} scan will find turning point $K$ (also on the shortest path), which if we can reach will get us to $J$. We shoot a ray to $K$ to find it again blocked by line segment $GH$, which when we recurse scan \CW{} will again find $H$, except this time in the check to bypass it, we have the angled sector $\AS{\Ptd{F}{K}}{\Ptd{F}{J}}$ and thus, trying to bypass will quickly leave the scan's angled sector along the line segment $HI$, therefore bypass has failed and we need to shoot to $H$.

\subsection{Caching Rays Shot}
The ray shots are the most expensive part of \orayscan{}, as  we will show in the experimental sections; therefore, reducing the cost of these shots can achieve great performance improvements. We make use of a simple method of remembering rays we shoot so that subsequent queries that shoot the same rays can just look at the result.

For our method for caching rays, we store the results of a ray shot from node $u$ to node $v$, where $u, v \notin \{s,t\}$.
The dynamic cases changes the results of stored ray, thus any changes in the environment invalidates the whole cache, and we start again.

\section{Extension to Multi-Target ESPP}
\rayscan{} can be extended to support single-source multi-target searches.
A naive way to do this is, during expansion, instead of looking for a single target within the projection field and shooting a ray, find all targets within the projection field and shoot rays to all.
A change of the heuristic function $\Hfn{u}$ either to $0$ or to the distance form $u$ to the closest target maintains the admissibility of the heuristic.

The number of additional rays shot to every target significantly increases runtime, which can be counteracted by the target blocking extension.

\begin{algorithm}[t]
\caption{Generate $u$'s successors within projection field}
\label{alg:successors}
	\begin{algorithmic}[1]
		\Function{start\_successors}{$s$, $T$} \label{lst:line:startSuccessors}
		    \State \Call{populate\_targets}{$T$, $\AS{\Ptd{s}{t}}{\Ptd{s}{t}}$} \label{lst:line:populateTargets}
			\State \Call{target\_successors}{$s$, $\AS{\Ptd{s}{t}}{\Ptd{s}{t}}$}\label{lst:line:s_targetSuccessors}
		\EndFunction
		
		\Function{successors}{$u$, $T$, $F=\AS{a_{\CCW}}{a_{\CW}}$} \label{lst:line:successors}
            \State \Call{populate\_targets}{$T$, $F$} \label{lst:line:populateTargets2}
			\State $(p,I,\_) \gets $ \Call{shootray}{$u$, $a_{\CCW}$} \label{lst:line:proFieldShootBegin}
			\State \Call{scan}{$u$, $p$, $I$, $F$, $T$, $\CW$}
		    \State $(p,I,R) \gets $ \Call{shootray}{$u$, $a_{\CW}$}
		    \If{$R$ is $\false$} \label{lst:line:refineRaySuccessors}
                \State \Call{scan}{$u$, $p$, $I$, $F$, $\CCW$} \label{lst:line:refineSectorSuccessors}
            \EndIf \label{lst:line:proFieldShootEnd}
			\State \Call{target\_successors}{$u$, $F$}\label{lst:line:Suc:targetSuccessors}
		\EndFunction
		
		\Function{target\_successors}{$u$, $F$} \label{lst:line:targetSuccessors}
			\While{$\Call{targets\_remaining}{} \ne \varnothing$} \label{lst:line:whileTargetsBegin}
			    \State $t \gets $ any target in \Call{targets\_remaining}{} \label{lst:line:anyTarget}
			    \State \Call{remove\_targets}{$t$} \label{lst:line:removeTarget}
				\State $(p,I,\_) \gets $ \Call{shootray}{$u$, $\Ptd{u}{t}$}\label{lst:line:TS:shoot}
				\If {$t$ is visible from $u$}
				    \State \Call{push\_successor}{$u$, $t$}\label{lst:line:TS:push}
				\Else
					\State \Call{scan}{$u$, $p$, $I$, $F$, $\CW$} \label{lst:line:refineSectorTS1}
					\State \Call{scan}{$u$, $p$, $I$, $F$, $\CCW$} \label{lst:line:refineSectorTS2}
				\EndIf
			\EndWhile \label{lst:line:whileTargetsEnd}
		\EndFunction
		
		\Function{shootray}{$u$, $a$} \label{lst:line:shootray}
		    \If{ray hits one or more turning points} \label{lst:line:rayTurnBegin}
    		    \State $n \gets$ first turning point $n$ hit
    		    \State \Call{push\_successor}{$u$, $n$} \label{lst:line:shootRayPush}
    		\EndIf \label{lst:line:rayTurnEnd}
    		\State Let $(p,I,R)$ be the first polygon $p$ blocking the ray\\
    		~~~~~~~~~~~~~~~at intersection $I$, and $R$ if shot to $a$ before
    	    \State \Return $(p,I,R)$
		\EndFunction
		
		\Function{scan}{$u$, $p$, $I$, $F$, $d$} \label{lst:line:scan}
            \State scan $p$ from $I$ in direction $d$ to find turning point $n$ \label{lst:line:scanSweep2}
            \State \Call{remove\_targets}{targets blocked during scan} \label{lst:line:sweepRemoveTargets}
            \If {scan leaves $F$} \Return \EndIf \label{lst:line:scanEnd1}
            \If {scan touches outer-boundary} \Return \EndIf \label{lst:line:scanEnd2}
            \State $(p',I',R) \gets $\Call{shootray}{$u$, $\Ptd{u}{n}$} \label{lst:line:scanShootRay}
            \If {$R$ is $\true$} \Return \EndIf \label{lst:line:scanEnd3}
            \If {$n$ is visible from $u$} \label{lst:line:scanVis}
                \If {$n$ is a forward turning point}
                    \State \Call{scan}{$u$, $p'$, $I'$, $\ASsplit{F}{\Ptd{u}{n}}{d}$, $d$} \label{lst:line:scanRecVis}
                \Else \Comment{$n$ is a backward turning point}
                    \State $d' \gets $ opposite orientation of $d$ \label{lst:line:scanRecVisBack1}
                    \State \Call{scan}{$u$, $p'$, $I'$, $\ASsplit{F}{\Ptd{u}{n}}{d'}$, $d'$} \label{lst:line:scanRecVisBack2}
                    \State \Call{scan}{$u$, $p$, $n$, $\ASsplit{F}{\Ptd{u}{n}}{d}$, $d$} \label{lst:line:scanRecVisBackF}
                \EndIf
	       \Else
                \State \Call{scan}{$u$, $p'$, $I'$, $\ASsplit{F}{\Ptd{u}{n}}{\CW{}}$, $\CW$} \label{lst:line:scanRecBlockCW}
                \State \Call{scan}{$u$, $p'$, $I'$, $\ASsplit{F}{\Ptd{u}{n}}{\CCW{}}$, $\CCW$} \label{lst:line:scanRecBlockCCW}
            \EndIf
		\EndFunction
	\end{algorithmic}
\end{algorithm}


Algorithm~\ref{alg:successors} details a successor generator for \rayscan{}, modified from \orayscan{} to support multiple targets. 
It can find paths from a start node $s$ to a list of target nodes $T$.

Function \Call{start\_successors}{$s$, $T$} (line~\ref{lst:line:startSuccessors}) will generate successors for source node $s$ that will lead to all targets $T$. Function \Call{successors}{$u$, $T$, $F$} similarly finds successors for the expanding node $u$ 
for all $t \in T$ within its projection field $F$.

Function \Call{populate\_targets}{$T$, $F$} (line~\ref{lst:line:populateTargets} and~\ref{lst:line:populateTargets2}) filters the list of targets $T$ that are within projection field $F$ and orders them circularly within the angled sector. These targets are stored within the \Call{targets\_remaining}{} global variable.

\begin{figure*}[t]
\centering
\begin{subfigure}[b]{0.31\textwidth}
\includegraphics[width=0.95\textwidth]{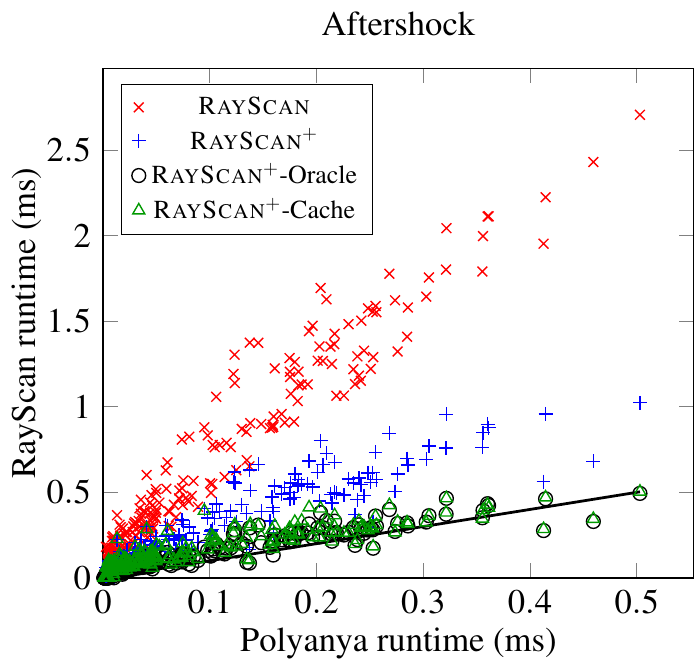}
\end{subfigure}
\begin{subfigure}[b]{0.3\textwidth}
\includegraphics[width=0.95\textwidth]{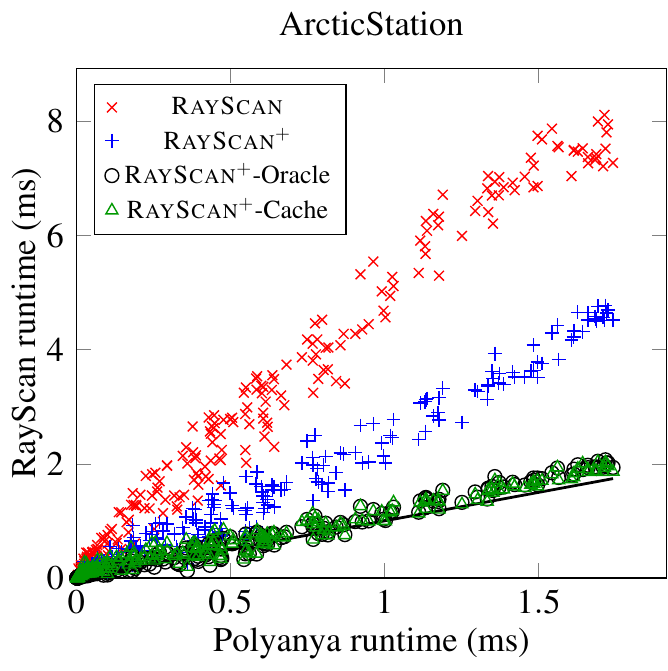}
\end{subfigure}
\begin{subfigure}[b]{0.31\textwidth}
\includegraphics[width=0.95\textwidth]{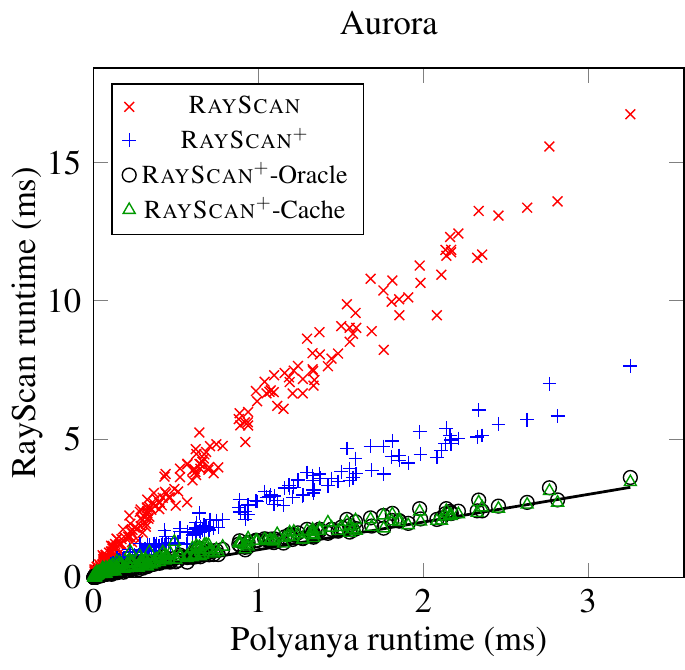}
\end{subfigure}
\caption{Scatter plots of Polyanya runtime vs \orayscan{} implementations for single target search}
\label{fig:rayscan-plus}
\end{figure*}

\begin{table}[tb]
\centering
\begin{tabular}{ | r | c | c | c | }
\hline
\textbf{Algs} & \textbf{Aftershock} & \textbf{Aurora} & \textbf{ArcticStation} \\ \hline
R & 568 & 3604 & 2752 \\ \hline
N & 400 & 2548 & 2689 \\ \hline
NB & 388 & 2492 & 2627 \\ \hline
NS & 375 & 2392 & 2090 \\ \hline
NP & 262 & 1638 & 1859 \\ \hline
NC & 150 & 736 & 683 \\ \hline
NBSP & 248 & 1560 & 1395 \\ \hline
NBSPC & 132 & 723 & 625 \\ \hline
\end{tabular}
\caption{Comparing separate extensions (rows) on different maps (columns); single-target average runtime per query (all times are in \SI{}{\micro\second}); (R) Original \orayscan{}; (N) convex scan \rayscan{}; (B) blocking extension; (S) skip extension; (P) bypass extension; and (C) ray caching}
\label{tab:ST}
\end{table}

Function \Call{shootray}{$u$, $a$} handles the ray shooting and returns the intersecting polygon $p$, the intersection point $I$, and a boolean $R$ indicating if shot in direction $a$ from $u$ has already been made this expansion. Lines~\ref{lst:line:rayTurnBegin} to~\ref{lst:line:rayTurnEnd} check for all collinear points the ray intersects with, and pushes the closest visible one as successor (if present) (line~\ref{lst:line:shootRayPush}). The ray will stop when it is blocked by an edge; if it only touches the corner of a polygon without entering (i.e. a collinear point) it will pass through it.

When expanding the start node $s$, we shoot to all the targets (line~\ref{lst:line:s_targetSuccessors}). The \Call{target\_successors}{$u$, $F$} (line~\ref{lst:line:targetSuccessors}) will handle all the target scans by shooting from the expanding node $u$ towards all remaining targets (when expanding node $u$ is the start node, all targets are the remaining targets).
The successors will continue to shoot to targets (lines~\ref{lst:line:whileTargetsBegin} to~\ref{lst:line:whileTargetsEnd}).
Specifically, it selects any of the remaining targets $t$ (line~\ref{lst:line:anyTarget}), removes it from the \Call{targets\_remaining}{} variable (line~\ref{lst:line:removeTarget}) and shoots towards it (line~\ref{lst:line:TS:shoot}).
If $t$ is visible, it is pushed as a successor (line~\ref{lst:line:TS:push}). Otherwise, the polygon that blocks the ray is scanned \CW{} and \CCW{}  while potentially removing additional targets that are found to be blocked (lines~\ref{lst:line:refineSectorTS1} and \ref{lst:line:refineSectorTS2}).

Expanding any node $u$ other than the start node will first do the projection scans by shooting along the extremes of its projection field $F$ and scan inwards of $F$ along the blocking polygon (lines~\ref{lst:line:proFieldShootBegin} to~\ref{lst:line:proFieldShootEnd}) like \orayscan{}.
Line~\ref{lst:line:refineRaySuccessors} is the refinement by ray.
These scans may have blocked some targets in \Call{targets\_remaining}{} which are removed.
The function \Call{target\_successors}{$u$, $F$} is called to shoot towards these remaining targets if any (line~\ref{lst:line:Suc:targetSuccessors}).

Function \Call{scan}{$u$, $p$, $I$, $F$, $d$}
is used to scan the polygon $p$ which blocked a ray shot from $u$ at the intersection point $I$. This function scans $p$ starting at $I$ in orientation $d$ (\CW{} or \CCW{})  while restricted to the angled sector $F$.
We first find a turning point by sweeping a line from $I$ in orientation $d$ to find a suitable turning point $n$ (line~\ref{lst:line:scanSweep2}). \rayscan{} uses convex scan to find a turning point whereas original \orayscan{} used forward scan.  This scan also uses the blocking, skipping and bypass extensions and removes targets that are found to be blocked during the scan (line~\ref{lst:line:sweepRemoveTargets}).

If the scan leaves the projection field $F$ or touches the outer-boundary, we terminate the scan  (lines~\ref{lst:line:scanEnd1} and~\ref{lst:line:scanEnd2}). Otherwise, we shoot a ray from $u$ towards $n$ (line~\ref{lst:line:scanShootRay}). Refine by ray is enforced on line~\ref{lst:line:scanEnd3} to terminate the scan if a ray towards $n$ was previously shot from $u$. If not, we recurse the scan, which differs slightly depending on whether $n$ is visible from $u$ (line~\ref{lst:line:scanVis})  or not. Like original \orayscan{}, if the visible turning point is a forward turning point, we recurse the scan in the same orientation with a split angled sector (line~\ref{lst:line:scanRecVis}). Otherwise, if it is a backward turning point, we recurse both ways, handling the opposite scan (lines~\ref{lst:line:scanRecVisBack1} and~\ref{lst:line:scanRecVisBack2}) and continuing to find the next forward turning point (line~\ref{lst:line:scanRecVisBackF}).
For a non-visible turning point, we follow the same principles as \orayscan{} in that we try to scan \CW{} around the blocking point $p$ (line~\ref{lst:line:scanRecBlockCW}) and \CCW{} (line~\ref{lst:line:scanRecBlockCCW}).

\section{Experiments}

The data set used for the experimentation stems from the Moving AI Lab pathfinding benchmarks \citep{sturtevant2012benchmarks}. 
We make use of three representative Starcraft maps: Aftershock, ArcticStation and Aurora, which are converted from the grid representation to Euclidean polygonal obstacles.

These experiments were conducted on a machine with an Intel i7-8750H, locked at \SI{2.2}{\giga\hertz} with boost disabled. Every search was run 7 times, discarding the best and worst results, then averaging the remaining 5. The source code will be made available\footnote{\texttt{https://bitbucket.org/ryanhech/rayscan/}}.

\begin{figure*}[t]
\centering
\begin{subfigure}[b]{0.31\textwidth}
\includegraphics[width=0.95\textwidth]{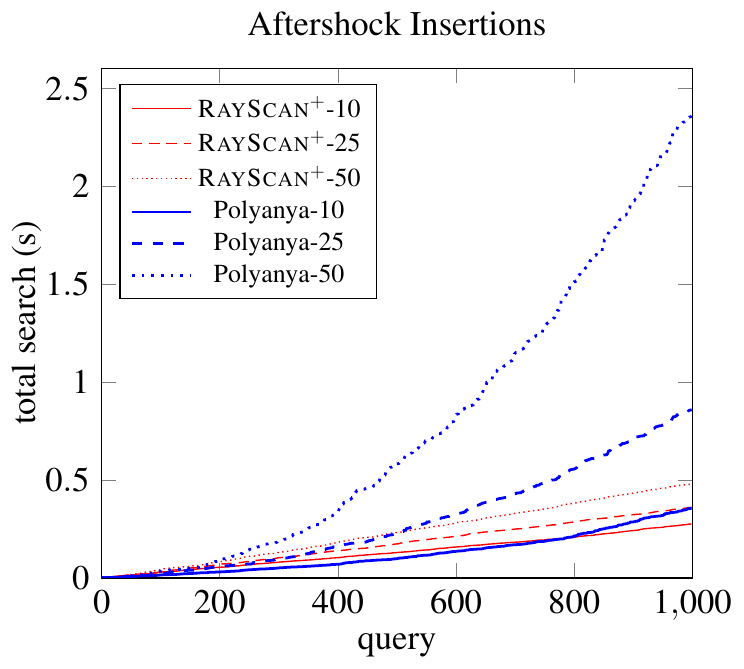}
\end{subfigure}
\begin{subfigure}[b]{0.3\textwidth}
\includegraphics[width=0.95\textwidth]{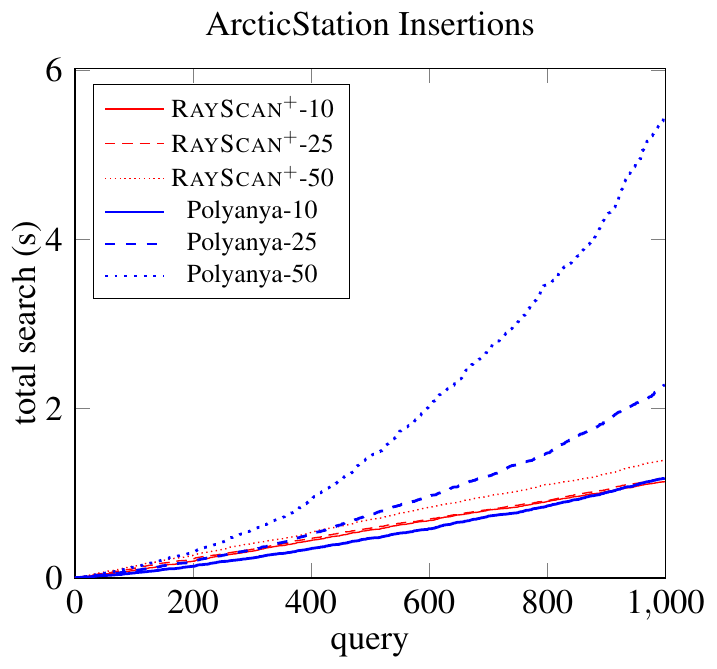}
\end{subfigure}
\begin{subfigure}[b]{0.3\textwidth}
\includegraphics[width=0.95\textwidth]{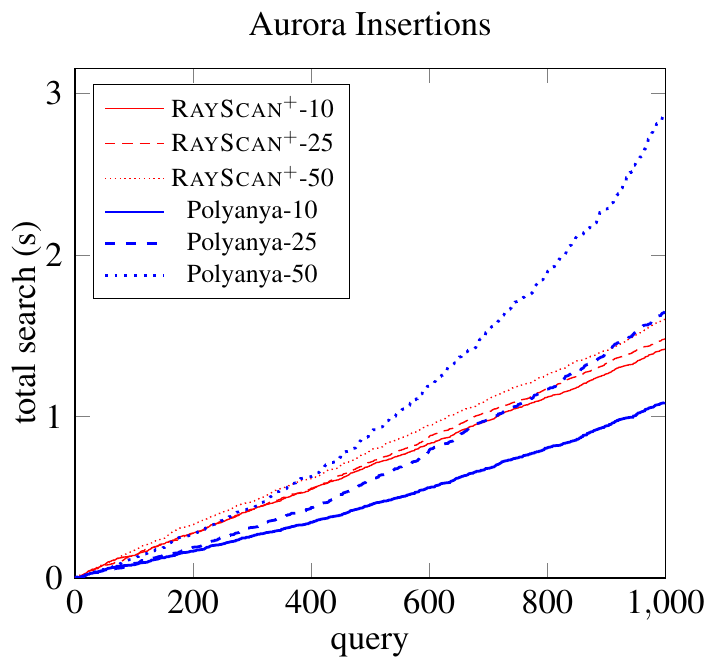}
\end{subfigure}
\caption{Dynamic accumulative search time (no setup) with varying number of inserts/removal of total obstacles every 10 single target instances}
\label{fig:dy-var}
\end{figure*}

\subsection{Comparison against Original RayScan}
The improvements listed in this paper can be seen in Figure~\ref{fig:rayscan-plus}, four different RayScan algorithms are compared against Polyanya. \orayscan{} shows the original RayScan implementation~\cite{rayscan}, which uses the forward-scan method and refinement by sector (although it was not detailed in the paper). \rayscan{} uses an updated implementation using the convex-scan method with target blocking, skip and bypass extensions and  refinement by ray. The \rayscan{}-Oracle is the same as \rayscan{} except the shooting of rays cost comes for free. \rayscan{}-Cache caches the results of ray shots from previous queries. The implementation of the ray shooting is done by drawing Bresenham's lines on a grid \citep{Bresenham}.

Our Polyanya implementation code is modified from the implementation by \citet{polyanya}. The navigation mesh was generated by constrained Delaunay triangulation (CDT) using Fade2D.\footnote{https://www.geom.at/products/fade2d/}
The triangle faces were greedily merged to make larger convex faces to improve Polyanya's performance.

Figure~\ref{fig:rayscan-plus} illustrates the significant advantages of \rayscan{} over \orayscan{} on single target problems.
These results show that Polyanya is still faster for searching than \rayscan{} for static environments, 
but adding caching to \rayscan{} achieves very competitive results. 
The disadvantage of caching rays is that for large maps this can lead to higher memory usage, as the number of rays shot can be in the order of the number of edges in the SVG. In the case where memory usage gets too large, caching strategies can be employed to ensure the memory usage never exceeds a given budget.

Table~\ref{tab:ST} provides an ablation study of the extensions we propose. Caching is clearly the most important extension (since ray shots are the most expensive part of the algorithm). Each other extension has a positive effect, with bypass the most effective (since the environments we test on 
have many non-convex polygons) and blocking the least (for single targets). The positive effects combine so best is using all of them. 

\subsection{Dynamic Single Target Scenarios}

We conduct experiments for dynamic environments with results shown in Figure~\ref{fig:dy-var}. 
These tests show the accumulated runtime ($y$-axis) of the algorithms over 1000 queries ($x$-axis), where we insert or remove 10, 25 or 50 small convex polygons in the environment every 10 queries, while leaving the original map untouched (i.e. no polygon originally on the maps is modified).

\ignore{
We see that Polyanya's first scenario starts at a much higher runtime, as it must first generate the navigation mesh, while \rayscan{} is much cheaper as it has to setup the nodes of the search (i.e. the vertices of the obstacles) and its ray shooter (i.e. drawing the polygons on a grid for Bresenham's), the cost of altering these structures for the inserted/removed obstacles are also accounted for every 10 queries that the environment is changed.
}

Inserting obstacles into Polyanya's mesh is done by splitting the faces the polygon lines intersect and constraining them.
To remove them, we clear the constraint between the faces along the obstacle edges. Polyanya also needs to find which face a point holds (for start, target and insert of new polygon). This is done by selecting a face and moving across the mesh in a direct line to the point.

\ignore{
Due to how we remove obstacles from the mesh, vertices of the mesh that are not adjacent to any edge constraints can be left behind, resulting in a more complicated mesh with more smaller faces present. While this makes it fast to remove obstacles from Polyanya, it does slow down Polyanya's search; to address this we added the Polyanya-Oracle, which includes the initial mesh generation cost, but for every change made a new mesh is generated for free to ensure the search is running on an optimal mesh.
}

 Figure~\ref{fig:dy-var} shows that \rayscan{}'s search performance is fairly consistent, as changes to its data structures are fast and do not degrade performance.
 Polyanya shows the performance is degrading as removing the obstacles is degrading the mesh; while this can be alleviated by repairing the mesh, this does incur additional costs in removal of unnecessary vertices and/or merging smaller faces together.

\ignore{
On a dynamic environment, changes to it results can invalidate of previous rays.
For simplicity, the entire cache is discarded when any changes are made (i.e. Figure~\ref{fig:dy-acc} \rayscan{}-Cache will empty the cache every 10 queries\footnote{Approaches to maintain unaffected cache entries could be used to substantially improve \rayscan{}-Cache}.)
}
\ignore{
Figure~\ref{fig:dy-var} shows us accumulated search times for both \rayscan{} and Polyanya, when comparing results where the rate of dynamic obstacle insertion/removal 
after each 10 queries is either 10, 20 and 50. 
\rayscan{} slowly degrades when the rate is higher as the search space is more complicated with additional obstacles present. 
But}
Polyanya's performance degrades significantly as the environment is changed more rapidly, because the navigation mesh gets more and more complicated.  Maintaining the mesh with methods by \citet{kallmann2004fully} that maintains the CDT or \citet{van2012navigation} that makes use of a Voronoi diagram become mandatory to maintain Polyanya's performance.
In highly dynamic maps \rayscan{} is faster. 

\begin{figure*}[tb]
\centering
\begin{subfigure}[b]{0.3\textwidth}
\includegraphics[width=0.95\textwidth]{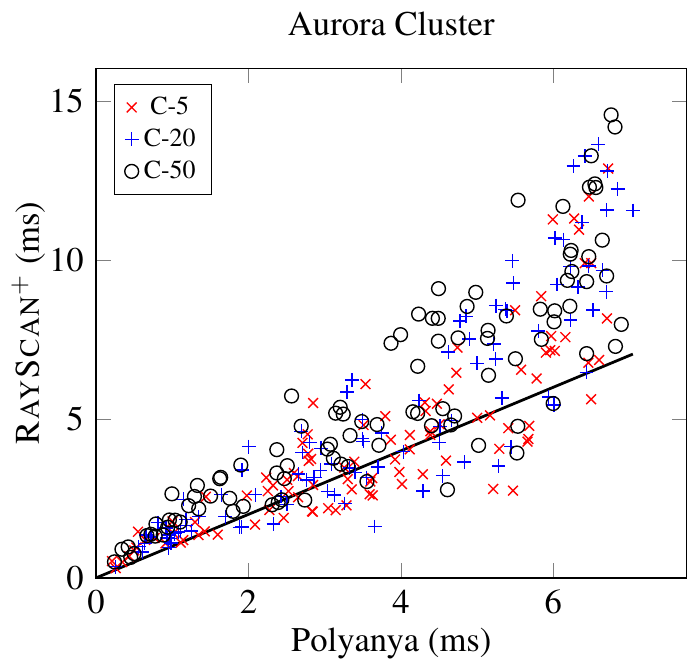}
\subcaption{\rayscan{} small targets}\label{fig:cluster-cmp}
\end{subfigure}
\begin{subfigure}[b]{0.29\textwidth}
\includegraphics[width=0.95\textwidth]{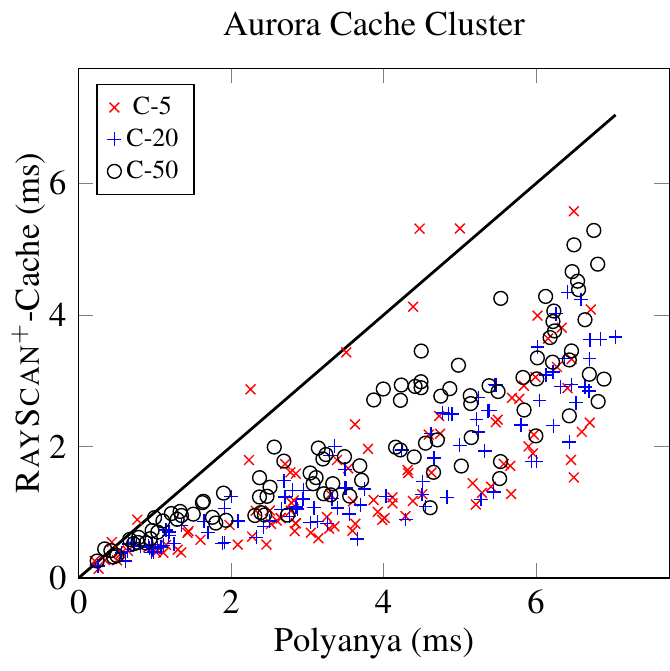}
\subcaption{\rayscan{}-Cache small targets}\label{fig:cluster-cmpc}
\end{subfigure}
\begin{subfigure}[b]{0.31\textwidth}
\includegraphics[width=0.95\textwidth]{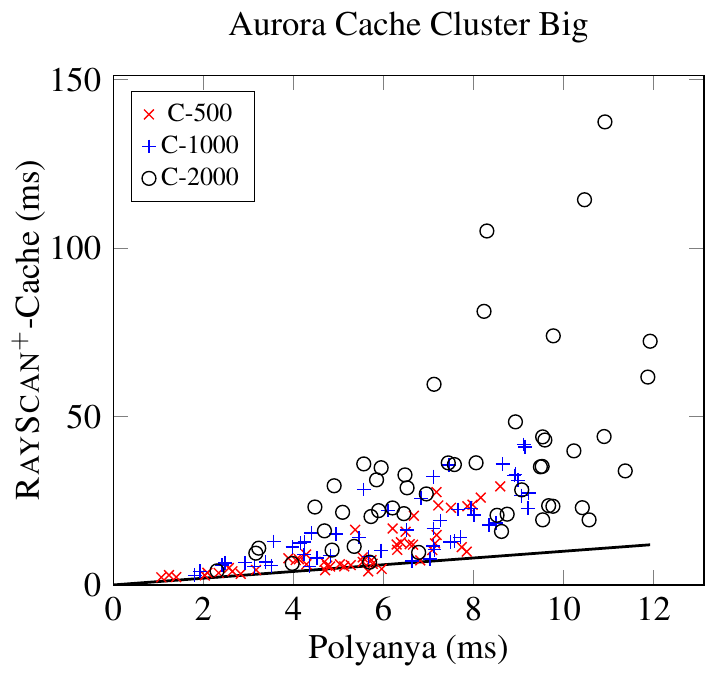}
\subcaption{\rayscan{} large targets}\label{fig:cluster-cmpc-large}
\end{subfigure} \\

\begin{subfigure}[b]{0.3\textwidth}
\includegraphics[width=0.95\textwidth]{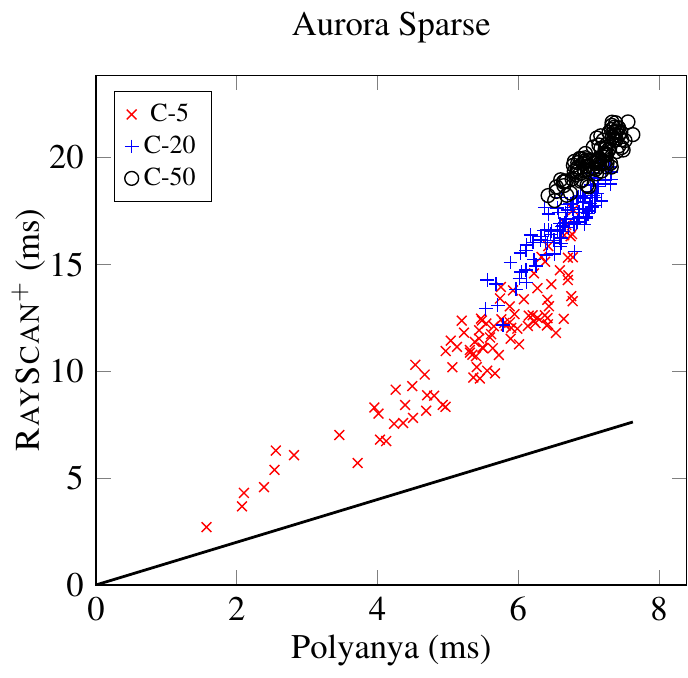}
\subcaption{\rayscan{} small targets}\label{fig:sparse-cmp}
\end{subfigure}
\begin{subfigure}[b]{0.3\textwidth}
\includegraphics[width=0.95\textwidth]{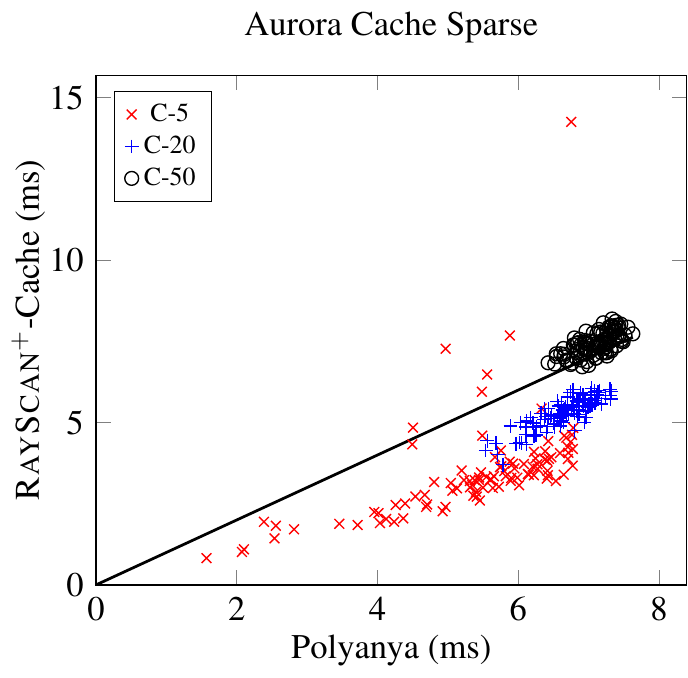}
\subcaption{\rayscan{}-Cache small targets}\label{fig:sparse-cmpc}
\end{subfigure}
\begin{subfigure}[b]{0.31\textwidth}
\includegraphics[width=0.95\textwidth]{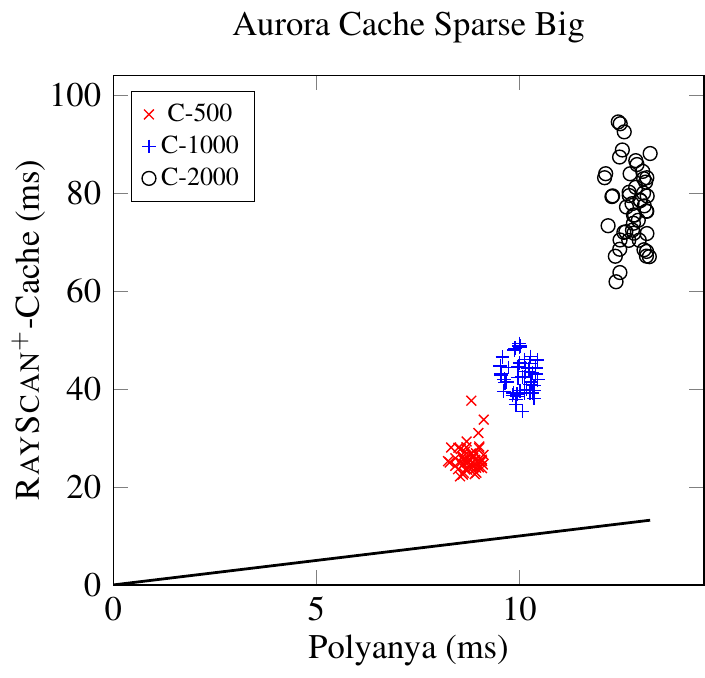}
\subcaption{\rayscan{} large targets}\label{fig:sparse-cmpc-large}
\end{subfigure}
\caption{Multi-target tests on map Aurora; (a)-(c) are with clustered targets; (d)-(f) are with sparse targets. C-$n$ details tests with $n$ number of targets of Polyanya ($x$-axis) vs \rayscan{} ($y$-axis)}\label{fig:sparse}
\end{figure*}

\subsection{Multi Target Scenarios}

In Figure~\ref{fig:sparse}a-c, we consider multi-target scenarios where the targets are clustered fairly close together. 
We have two types of scenarios: Figure~\ref{fig:cluster-cmp} and~\ref{fig:cluster-cmpc} split the Aurora map into 10x10 grid cells and randomly choose a number of different targets within one of those cells. Figure~\ref{fig:cluster-cmpc-large} evaluates the impact of a larger number of targets which are placed in a randomly chosen cell from a 6x6 grid.

Multi-target Polyanya makes use of the interval heuristic \citep{zhao2018fast}, which produces a Dijkstra like expansion for Polyanya. This can impact the search performance in clustered examples 
since Polyanya is not actively seeking out the targets; however, Polyanya only needs to consider each target during the search when it reaches a face containing the target.

Using \rayscan{} for multiple targets makes heavy use of the blocking extension, as otherwise we need to shoot to all targets within the projection field of each node expanded, resulting in many additional rays. 
\rayscan{} uses the Euclidean distance to the closest target to the vertex. 
For a small number of targets (e.g. Figure~\ref{fig:cluster-cmp}), the heuristic and selection of targets within each projection field is done by checking all targets. For larger number of targets (e.g. Figure~\ref{fig:cluster-cmpc-large}) we use a Hilbert R-Tree \citep{kamel1993hilbert} to speed up these calculations.

Examining Figure~\ref{fig:cluster-cmp}, we see that for clustered targets Polyanya is fairly consistent as the number of targets change, compared to \rayscan{}, which slows down as the number of targets grow. This is expected as Polyanya only needs to handle targets at the beginning (to locate each target's face) and the end (to finalise the path to target); whereas \rayscan{} needs to deduce for each node expansion which targets are in its projection field.

Figure~\ref{fig:cluster-cmpc} illustrates that when caching the rays, \rayscan{} is competitive with Polyanya. It is faster than Polyanya when the targets are clustered together. 
The high spikes for C-5 are early searches still building up the cache.

Figure~\ref{fig:cluster-cmpc-large} compares results for a larger number of targets. 
This highlights a weakness of \rayscan{} since it has to consider each target within its projection field for every expansion.
\rayscan{} must do this as it only produces a subset of successors, which means vital successors needed to reach a target might not be found without shooting to that target from certain nodes. Until a method of addressing this weakness is found, \rayscan{} will struggle with very large numbers of targets compared to Polyanya.

Figure~\ref{fig:sparse}d-f compares the methods for 
sparse target points, which are distributed at random around the map.
The nearest point heuristic for \rayscan{} is less effective in these scenarios, resulting in \rayscan{} performance degrading more with each additional point, as the Polyanya interval heuristic is more useful in these cases. 
We see that the \rayscan{}-Cache speeds perform slightly worse at 50 targets, but is highly competitive with fewer targets.

An ablation study shows that for multi-target blocking is an important extension: for 5-50 targets it leads to 20\% improvements; for 500-2000 it can speed up by more than 2$\times$.

\ignore{
\begin{table}[tb]
\centering
\begin{tabular}{ | r | c | c | }
\hline
\textbf{Algs} & \textbf{Aurora Sparse} & \textbf{Aurora Sparse Big} \\ \hline
N & 18.83k & 231k \\ \hline
NB & 16.02k & 99.42k \\ \hline
RB & 16.44k & 63.98k \\ \hline
\end{tabular}
\caption{Comparing extension for multi-target search, refer to Table~\ref{tab:ST} captions for details}\label{tab:MT}
\end{table}

Table~\ref{tab:MT} shows the ablation study for 
multi-target search, the striking difference here with Table~\ref{tab:ST} is the effectiveness of blocking. 
}

\section{Conclusion}
\rayscan{} is an efficient method for Euclidean shortest path finding in dynamic situations, since it requires almost no pre-processing to run.  In this paper we show how to substantially improve \rayscan{} by reversing the order of target and projection scans, and reducing the number of vertices we need to shoot rays at. We extend \rayscan{} to shoot at multiple targets.  
\rayscan{} is competitive with the state of the art ESPP method Polyanya when we cache ray shots, which make up the principle cost of \rayscan{}.  Future work will examine better methods to maintain ray shot caching in dynamic situations.

\section{Acknowledgements}
Research at Monash University is supported by the Australian Research Council (ARC) under grant numbers DP190100013, DP200100025 and FT180100140 as well as a gift from Amazon.

\bibliography{main}
\end{document}